\documentclass[reprint,pre,amsmath,amssymb]{revtex4-1}
\usepackage{graphicx}
\usepackage{color}
\usepackage{mathtools}
\usepackage{bm}
\usepackage{comment}
\usepackage{lineno,hyperref}

\usepackage[export]{adjustbox}

\usepackage{float}

\usepackage{subfigure}
\usepackage{dcolumn}
\newcolumntype{x}[1]{>{\centering\arraybackslash\hspace{0pt}}p{#1}}
\modulolinenumbers[1]

\usepackage[colorinlistoftodos,prependcaption,textsize=tiny]{todonotes}
\usepackage{soul}

\begin{document}

\begin{abstract}
Extrusion is a widely used process for forming pastes into designed shapes, and is central to the manufacture of many industrial products. The extrusion through a square-entry die of a model paste of non-Brownian spheres suspended in a Newtonian fluid is investigated using discrete element simulations, capturing individual particle contacts and hydrodynamic interactions. The simulations reveal inhomogeneous velocity and stress distributions, originating in the inherent microstructure formed by the constituent particles. Such features are shown to be relevant to \emph{generic} paste extrusion behaviour, such as die swell. The pressure drop across the extruder is correlated with the extrudate velocity using the Benbow-Bridgwater equation, with the empirical parameters being linked directly to particle properties such as surface friction, and processing conditions such as extruder wall roughness. Our model and results bring recent advances in suspension rheology into an industrial setting, laying foundations for future model development, paste formulation and extrusion design.
\end{abstract}

\title{Linking particle properties to paste extrusion flow\\characteristics using discrete element simulations}
\author{Christopher Ness}
\affiliation{School of Engineering, University of Edinburgh, Edinburgh EH9 3JL, United Kingdom}%
\author{Jin Y. Ooi}
\affiliation{School of Engineering, University of Edinburgh, Edinburgh EH9 3JL, United Kingdom}%
\author{Jin Sun}
\affiliation{School of Engineering, University of Edinburgh, Edinburgh EH9 3JL, United Kingdom}%
\author{Michele Marigo}
\affiliation{Johnson Matthey Technology Centre, Billingham, PO Box 1, TS23 1LB, United Kingdom}
\author{Paul McGuire}
\affiliation{Johnson Matthey Technology Centre, Billingham, PO Box 1, TS23 1LB, United Kingdom}
\author{Han Xu}
\affiliation{Johnson Matthey Technology Centre, Billingham, PO Box 1, TS23 1LB, United Kingdom}
\author{Hugh Stitt}
\affiliation{Johnson Matthey Technology Centre, Billingham, PO Box 1, TS23 1LB, United Kingdom}
\date{\today}
\maketitle

\section{Introduction}
Pastes are highly concentrated dispersions of solid particles in a liquid. Extrusion is a ubiquitous process for forming such pastes into complex predetermined shapes, and is a crucial step in the manufacture of many consumer and industrial products, such as porcelains and catalyst supports \cite{Benbow1993}. To meet increasing technical demands on finished product structure and performance, improvements to existing manufacturing methods are urgently required. Minor changes to formulation or operating conditions can, however, result in flow instabilities and product defects that are very challenging to predict with the models currently available~\cite{Green1953,Benbow1987,Horrobin1998,Draper1999}.

During extrusion, the paste is forced to flow through a constriction with inlet/outlet cross-sectional area ratio $A_0/A$, and along a narrow die of length $L$ and perimeter $M$, giving a product of desired composition and dimensions.
The classical expression used to relate the pressure drop $\Delta P$ required to achieve an extrudate velocity $V$ is given by Benbow and Bridgwater~\cite{Benbow1993} as
\begin{equation}
\Delta P=(\sigma_0 + \alpha V^m) \ln \left( \frac{A_0}{A}  \right) + (\tau_0+\beta V^n)\frac{ML}{A} \text{,}
\label{eq:1}
\end{equation}
where $\alpha$, $\beta$, $m$, $n$, $\sigma_0$, $\tau_0$ are related to the bulk constitutive material properties and processing conditions, and can be calibrated using experimental data \cite{Wilson2006}. This semi-empirical expression has become popular for both industrial and academic extrusion pressure calculations~\cite{Martin2002}. It relies on a continuum description of the material rheology, assuming the paste can be characterised simply by a bulk yield stress $\sigma_0$ and a zero-velocity wall shear stress $\tau_0$ in addition to power-law shear-rate dependence~\cite{Zheng1992}.

In reality, the pastes being extruded are disordered, amorphous, solid-liquid assemblies~\cite{Sollich1997} whose rheology typically exhibits many non-linearities such as yield stress behaviour~\cite{Pham2007}, shear thinning~\cite{Guy2015} and shear thickening~\cite{Brown2010a,Brown2009,Bertrand2002,Wagner2009}, as well as time-dependence~\cite{Lapasin1983}. These phenomena derive at the particle level and are governed by details of the particle-scale physics as demonstrated recently for the case of shear thickening in a series of experimental and numerical works (see, for example~\cite{seto2013discontinuous,Fernandez2013,Lin2015}), and traditional continuum rheological models, for example Bingham or Herschel-Bulkley, cannot provide a \textit{comprehensive} description of the rheology across flow regimes~\cite{Zhou2013}. Moreover, in extreme cases the flow confinement may be such that separation of scales between the die geometry $A$ and a typical particle diameter $d$ is only one order of magnitude or less, rendering the dynamics more comparable to, for example, discharging grains~\cite{Holst1999,Wang2015}, than to those of a continuous non-Newtonian fluid. In such scenarios, the discrete nature of the paste is of paramount importance to the rheology.
A further complicating flow phenomenon of particular interest is liquid phase migration (LPM)~\cite{Khelifi2013,Rough2004}, the net movement of liquid down an internal pressure gradient. LPM results in localised fluctuations in the solids content of the paste, leading to regions of increased viscosity~\cite{Andreotti2012,Suzuki2015} or undesired plastic or quasi-static behaviour in an otherwise viscous flow \cite{Yu1999, Rough2000,Elhweg2009}, rendering a continuum description of the material highly challenging. Several studies of ram extrusion have found strong evidence that the onset of LPM, proven to be present both by post-extrusion drying, weighing and analysis of extrudate samples, and by ram force-time profiles, is correlated with ram velocity \cite{Rough2004, Mascia2006, Zhang2011, Hongjun2009}. It is more likely to occur at low velocities \cite{Rough2004, Hongjun2009}, where the convection of liquid through the slower moving solid matrix is more prevalent. Furthermore, it has been proposed that in addition to localised pressure gradients, LPM during extrusion can also be caused by suction effects due to the extensional strain rate imposed on the paste at the die entry \cite{Mascia2006, Patel2007, Zhang2011}.

As a result of these complications, Equation~\ref{eq:1} may, if suitably calibrated, offer a practical determination of the energy input required for specific extrusion processes, but it does not have predictive capability for new materials whose constitutive behaviour is unknown, or not describable simply as Herschel-Bulkley or Bingham.
It is clear, therefore, that new constitutive equations relating particle properties and particle-scale structural and stress information to processing parameters~\cite{Coussot2007,Sun2011} are required to provide thorough and predictive alternative models for extrusion, in order to meet the future challenges in paste formulation and process design.

A number of semi-empirical constitutive models have been proposed to address this issue specifically for extrusion flow in industrial and consumer materials \cite{Aydin2000,Peck2006}, using experimentally observed flow features to build upon the traditional constitutive equations~\cite{Roussel2005}. Such models typically correlate temporal microstructural changes during extrusion to bulk rheological properties. For example, Engmann and Mackley~\cite{Engmann2006} modelled chocolate during cold extrusion as a Bingham plastic by dynamically relating the yield stress to the energy required to convert the constituent crystalline particles to a liquid state. Similarly, the Carreau Model, a constitutive equation typically used for describing the rheology of polymers, has been adapted~\cite{Patil2008} to describe the microstructural evolution of PTFE paste during extrusion. This allows the model to begin to predict shear-thickening behaviour as the microstructure develops during extrusion, giving remarkable agreement with experimental data. 

In the present work, we address this need for a stronger rational link between particle-scale physics and process-level extrusion rheology.
Discrete element method (DEM) simulations are carried out, resolving the velocities and contacts of all suspended particles using an interaction model that has been demonstrated to successfully capture the shear rheology of suspensions~\cite{Ness2015a,Ness2016b} with convincing experimental agreement.
We investigate pressure-driven extrusion flow through a square-entry die, and present the resulting internal velocity and stress profiles predicted by the model. These profiles provide a general description of extrusion flow governed at the particle-, rather than bulk-, scale, and may serve as a template for future constitutive descriptions of paste rheology.
We further elucidate the links between the observed bulk extrusion behaviour and the particle-scale properties by describing the flow according to the Benbow-Bridgwater equation and explicitly considering the relationships between particle stiffness, particle friction, liquid viscosity and wall roughness and the Benbow-Bridgwater parameters.
Overall, our results highlight the discrete nature of the paste and allow us to shed light on the links between microscale physics and macroscale extrusion phenomena, providing a foundational basis for the development of future constitutive equations as well as guiding future practice for industrial paste formulation and process design. Our model and the present results are relevant to processing of soft matter in general, and may have broad impact and utility across industry.

\section{Description of Simulation Model}
\subsection{Contact model for particle and hydrodynamic interactions}
\label{sec:contact_model}
We consider a model paste composed of purely repulsive spherical particles, density ($\rho$) matched with an interstitial liquid of viscosity $\eta_f$. The paste is bidisperse, being composed of a 1:1 mixture, by number, of particles with diameters $d$ and $1.5d$. Such a particle size distribution is adopted to prevent crystallisation as occurs for monodisperse particles \cite{Yunker2010,Ikeda2012} .
The paste is assumed to be non-Brownian, meaning particle thermal diffusion is negligible when compared to advection under the driving flow. This non-Brownian limit, corresponding to P\'{e}clet number P\'{e} $( = 3\pi \eta_f \dot{\gamma}d^3/4kT)\gg1$ for characteristic advection rate $\dot{\gamma}$ and thermal energy $kT$, arises in dense suspensions of both silica and polymethylmethacrylate, for example, under typical processing conditions~\cite{Lin2015}.
The equation of motion for the suspended particles can be written simply as \cite{Brady1988}
\begin{equation}
m \frac{d}{dt}\binom{\mathbf{v}}{\boldsymbol{\omega}} = \sum \binom{\mathbf{F}}{\mathbf{\Gamma}} \text{,}
\label{eq:newton}
\end{equation}
for particles of mass $m$ with translational and rotational velocity vectors $\mathbf{v}$ and $\boldsymbol{\omega}$ respectively, subjected to force and torque vectors $\mathbf{F}$ and $\mathbf{\Gamma}$ respectively. For the model paste studied in this work, forces and torques arising due to direct contacts between neighbouring particles ($\mathbf{F}^c, \mathbf{\Gamma}^c$) and through particle-fluid hydrodynamic interactions ($\mathbf{F}^h, \mathbf{\Gamma}^h$) are considered.
Full solution of the hydrodynamic forces has traditionally been achieved computationally using the Stokesian Dynamics algorithm \cite{Bossis1984, Brady1985,Brady2001}, which resolves long-range, many-body interactions in addition to short-range pairwise, divergent lubrication films. The great computational expense of such an approach makes large simulations ($> 3000$ particles) challenging. For very dense suspensions, for which the mean neighbouring-particle separation becomes extremely small relative to the particle diameter, the divergent lubrication resistances dominate the hydrodynamic interaction \cite{Ball1997}, while long-range hydrodynamic interactions are screened by the presence of numerous intervening particles. $\mathbf{F}^h$ and $\mathbf{\Gamma}^h$ can therefore be approximated by summing pairwise lubrication forces among neighbouring particles (\cite{Ball1997,Kumar2010, Trulsson2012,Mari2014,Ness2015a} and others). For an interaction between particles $i$ and $j$, the force and torque on particle $i$ can be expressed as
\begin{subequations}
\begin{align}
&
\begin{multlined}
\mathbf{F}^{h}_{ij} = -a_{sq} 6 \pi \eta_f (\textbf{v}_i - \textbf{v}_j) \cdot \textbf{n}_{ij} \textbf{n}_{ij}\\
- a_{sh} 6 \pi \eta_f (\textbf{v}_i - \textbf{v}_j) \cdot (\mathbf{I}-\mathbf{n}_{ij}\mathbf{n}_{ij}) \text{,}
\end{multlined}\\
&
\begin{multlined}
\mathbf{\Gamma}^{h}_{ij} = -a_{pu} \pi \eta_f d_i^3 (\boldsymbol{\omega}_i - \boldsymbol{\omega}_j) \cdot (\mathbf{I}-\mathbf{n}_{ij}\mathbf{n}_{ij})\\
- \frac{d_i}{2} \left(\mathbf{n}_{ij} \times \mathbf{F}^{l}_{i}\right) \text{.}
\end{multlined}
\end{align}
\label{eq:lube_forces}
\end{subequations}
for particle diameter $d_i$, centre-to-centre unit vector from $j$ to $i$ $\mathbf{n}_{ij}$ and identity tensor $\mathbf{I}$. The squeeze $a_{sq}$, shear $a_{sh}$ and pump $a_{pu}$ resistance terms, derived by \cite{Kim1991} for $\beta = d_j/d_i$, are given by

\begin{subequations}
\begin{align}
&
\begin{multlined}
a_{sq} = \frac{\beta^2}{(1+\beta)^2} \frac{d_i^2}{4h_\text{eff}}	+	\frac{1 + 7\beta + \beta^2}{5(1 + \beta)^3} \frac{d_i}{2} \ln \left(\frac{d_i}{2h_\text{eff}} \right)\\
+ \frac{1 + 18\beta - 29\beta^2 + 18\beta^3 + \beta^4}{21(1+\beta)^4} \frac{d_i^2}{4h_\text{eff}} \ln \left(\frac{d_i}{2h_\text{eff}}\right) \text{,}
\end{multlined}\\
&
\begin{multlined}
a_{sh} = 4 \beta \frac{2 + \beta + 2\beta^2}{15 (1 + \beta)^3} \frac{d_i}{2} \ln \left( \frac{d_i}{2h_\text{eff}}\right)\\
+ 4\frac{16 -45\beta + 58\beta^2 - 45\beta^3 + 16\beta^4}{375(1 + \beta)^4} \frac{d_i^2}{4h_\text{eff}} \ln \left( \frac{d_i}{2h_\text{eff}} \right) \text{,}
\end{multlined}\\
&
\begin{multlined}
a_{pu} = \beta \frac{4 + \beta}{10(1 + \beta)^2} \ln \left( \frac{d_i}{2h_\text{eff}} \right)\\
+ \frac{32 - 33\beta + 83\beta^2 + 43\beta^3}{250(1+\beta)^3} \frac{d_i}{2h_\text{eff}} \ln \left( \frac{d_i}{2h_\text{eff}} \right) \text{.}
\end{multlined}
\end{align}
\label{eq:resistances}
\end{subequations}
For each pairwise interaction, the interparticle gap, i.e. the surface-to-surface distance $h$, is calculated according to $h = |\mathbf{r}_{ij}| - \frac{d_i + d_j}{2}$ for centre-to-centre vector $\mathbf{r}_{ij}$.
An increasing body of experimental work \cite{Fernandez2013,Guy2015} indicates that direct particle-particle surface contacts can play a major role in determining paste viscosity; indeed, simulations that strictly resolve hydrodynamic forces, treating particles as ideally hard and ideally smooth~\cite{Bossis1989}, have proven to be inadequate for capturing dense suspension rheology for cases where particle-particle contacts are presumed to be important. Moreover, even simulations that capture the elasto-hydrodynamic deformation of particles under large lubrication forces~\cite{Jamali2015} have proven insufficient to capture the large viscosities of suspended frictional particles. We therefore truncate the lubrication divergence and regularize the contact singularity at a typical asperity length scale $h_\text{min}=0.001d_{ij}$ for weighted average particle diameter $d_{ij} = \frac{d_id_j}{d_i + d_j}$, i.e., setting $ h = h_\text{min}$ in the force calculation, when $h < h_\text{min}$, allowing particles to come into contact.
The effective interparticle gap used in the force calculation, $h_\text{eff}$, is therefore given by
\begin{equation}
    h_\text{eff} =  \left\{
                \begin{array}{ll}
                  h & \text{for } h > h_\text{min}\\
                  h_\text{min} & \text{otherwise.} 
                \end{array}
              \right.
\end{equation}
For computational efficiency, the lubrication forces are omitted when the interparticle gap $h$ is greater than $h_\text{max} = 0.05 d_{ij}$. The volume fraction is sufficiently high that all particles have numerous neighbours within this range, so such an omission is inconsequential to the dynamics, as we verified elsewhere~\cite{Ness2015a}.
When a lubrication film between two particle surfaces ruptures (this occurs when $h < 0$), 
the particle interaction is defined according to a purely repulsive linear spring model \cite{Cundall1979}, with normal $\mathbf{F}^{c,n}$ and tangential $\mathbf{F}^{c,t}$ forces and torque $\mathbf{\Gamma}^c$ given by
\begin{subequations}
\begin{equation}
\mathbf{F}^{c, n}_{ij} = k_\text{n} \delta \mathbf{n}_\text{ij} \text{,}
\end{equation}
\begin{equation}
\mathbf{F}^{c,t}_{ij} = -k_\text{t} \mathbf{u}_\text{ij} \text{,}
\end{equation}
\begin{equation}
\mathbf{\Gamma}^{c}_{i} = -\frac{d_i}{2} (\mathbf{n}_{ij} \times \mathbf{F}^{c,t}_{i,j}) \text{,}
\end{equation}
\end{subequations}
for a collision between particles $i$ and $j$ with normal and tangential spring stiffnesses $k_n$ and $k_t$ respectively, particle overlap $\delta$ and tangential displacement $\mathbf{u}_\text{ij}$, updated incrementally during each contact. We note that the damping arising from hydrodynamic interactions is always sufficient to achieve a steady state, so further damping in the particle-particle contact model, a typical feature of simulations of dry granular matter~\cite{Herrmann1998}, is omitted for simplicity. A particle-particle Coulomb friction coefficient $\mu_p$ is defined according to  $|\mathbf{F}^{c,t}_{i,j}| \leq \mu_p|\mathbf{F}^{c,n}_{i,j}| $, setting a maximal value for the tangential force exerted during a collision. This friction coefficient sets a critical force value for which contacting particles will slide past each other rather than roll over each other. In addition, a particle-wall friction coefficient $\mu_w$ is defined, regulating the maximal tangential force that may be exerted by the wall on any particle.

\begin{figure*}
  \centering
 \includegraphics[trim = 5mm 111mm 10mm 1mm, clip,width=0.85\textwidth]{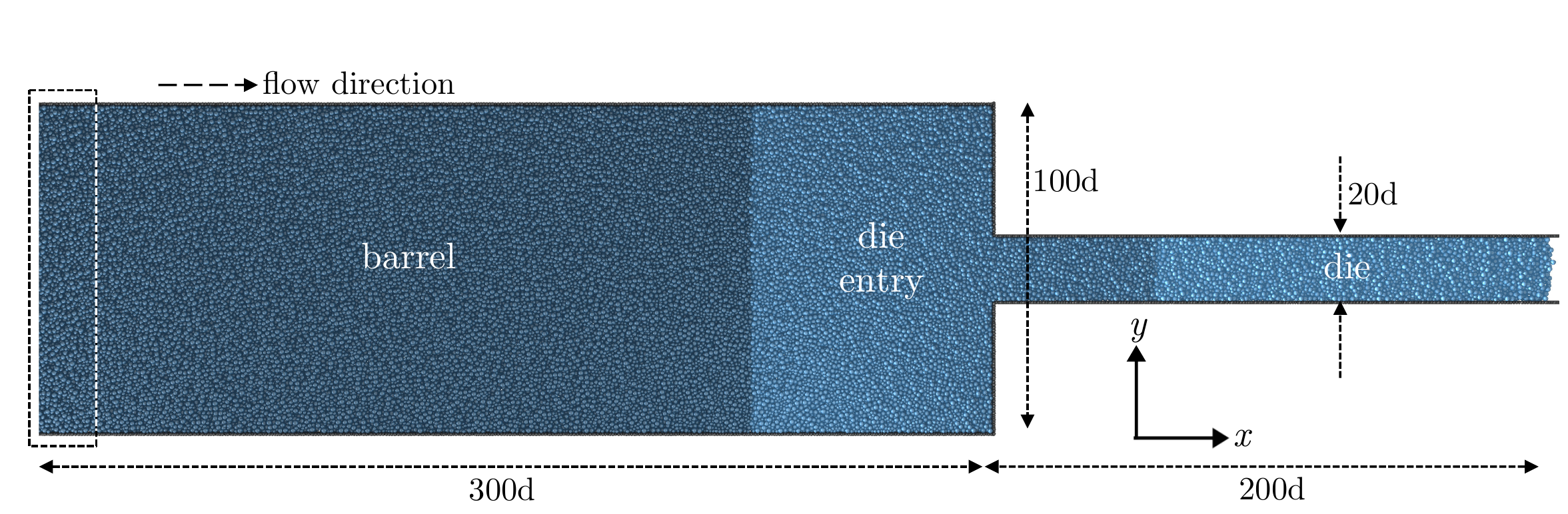}
     \caption{Schematic of the extrusion geometry showing barrel, die entry and die as represented by different particle shadings, with flow direction from left to right as indicated. Dashed arrows indicate the dimensions in units of particle diameters $d$; solid arrows indicate the coordinate definition. The geometry is $5d$ deep in the $z$-direction, with a periodic boundary condition on front and back. The dashed box at the far left illustrates the particle insertion and force application region.}
 \label{fig:geometry1}
\end{figure*}

\begin{figure}
  \centering
\includegraphics[width=0.4\textwidth]{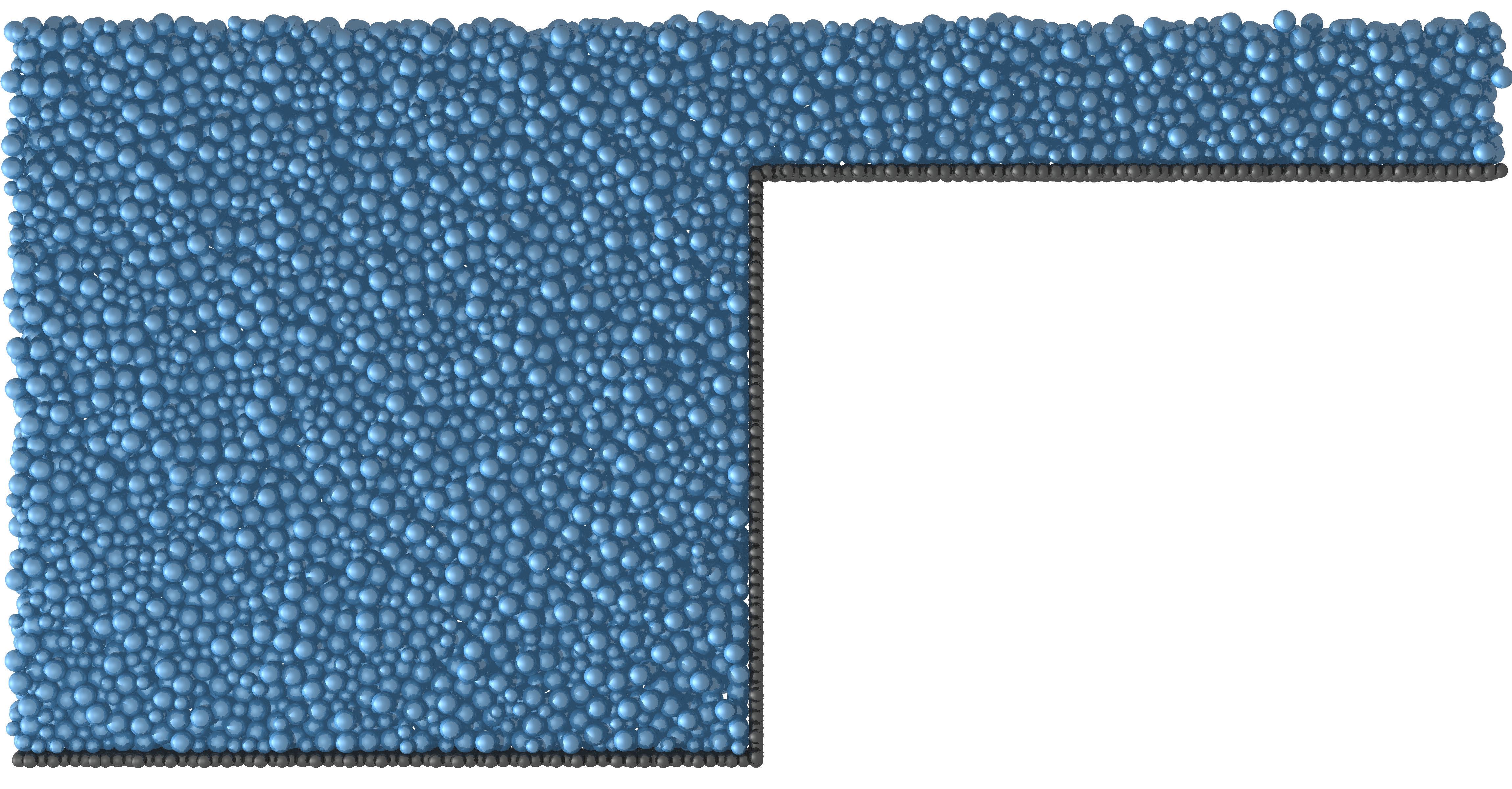}
     \caption{Close-up image of die entry region, to illustrate the bounding walls composed of nonmoving, smaller particles.}
 \label{fig:geometry2}
\end{figure}

The trajectory of each particle is calculated based on Equation~\ref{eq:newton}, where $\Sigma \mathbf{F}$ and $\Sigma \bm{\Gamma}$ represent the sums of all particle-particle and particle-fluid forces and torques acting on that particle. The integration is carried out in a step wise, deterministic manner, employing the Velocity-Verlet algorithm to update particle positions and velocities, implemented in LAMMPS~\cite{Plimpton1995}. Further justification of the model features and assumptions is given elsewhere~\cite{Ness2016a}.

Per-particle hydrodynamic and contact stresslet (force dipole) contributions are calculated according to
\begin{subequations}
\begin{align}
\bm{\Sigma}^H_i &=   \sum_{i \neq j} \mathbf{r}_{ij} \otimes \mathbf{F}^h_{ij} \text{,} 
\label{eq:stressF} \\
\bm{\Sigma}^C_i &=   \sum_{i \neq j} \mathbf{r}_{ij} \otimes \mathbf{F}^c_{ij} \text{,}
\label{eq:stressC}
\end{align}
\end{subequations}
respectively \cite{Ness2015a}, where $\mathbf{F}^c_{ij} = \mathbf{F}^{c,n}_{ij} + \mathbf{F}^{c,t}_{ij}$. The stresses that we consider in this work are derived from the total stresslet $\bm{\Sigma}_i=\bm{\Sigma}^H_i + \bm{\Sigma}^C_i$~\cite{Gallier2014} as described below. According to this sign convention, compressive stresses are \textit{positive}, consistent with the granular mechanics community, but opposite to that typically used in the rheology community.

\subsection{Geometry and boundary conditions}

A model paste described by the above set of interaction forces is extruded through a pseudo-2D square entry extrusion die with geometry and dimensions given in Figure~\ref{fig:geometry1}. A barrel width of $100d$ and a die width of $20d$ are used to approximate very narrow confinement such as might arise in an industrial extrusion setting, for example the extrusion geometry pertaining to the inner honeycomb walls of a monolithic ceramic catalyst support. While the die width of $20d$ may be realistic for the most narrow of industrially employed extrusion dies, the corresponding barrel width in such an apparatus would generally be considerably larger than $100d$. The present dimensions are chosen, however, to reduce computational expense while retaining relevant features of the geometry such that the resulting velocity and stress profiles may be indicative of those in reality. The geometry is periodic in the $z$-direction, with a depth of $5d$.

The extruder walls are composed of particles of diameter $0.5d$ at very high area fraction, giving an inherent wall bumpiness that will offer some resistance to flow by allowing the walls to support non-zero loads in the $x$-direction, Figure~\ref{fig:geometry2}. Such an approach has been adopted previously to study lubricated suspension flows under confinement~\cite{Nott1994,Singh2000,Chow2015}. An identical contact model to that presented in Section~\ref{sec:contact_model}, comprising hydrodynamic and contact forces, describes the interaction between paste particles and wall particles, though a separate friction coefficient $\mu_w$ is defined independently of $\mu_p$, allowing the frictional behaviour between particles and wall to be different to that for particle-particle contacts. Interactions between neighbouring constituent wall particles are excluded from the numerical integration, effectively freezing those particles in their pre-determined positions.

A pressure drop along the axial length of the extruder is generated by applying a constant body force $F$ in the $+x$ direction at each timestep to all particles that are inside the dashed box in Figure~\ref{fig:geometry1}. These particles will consequently move downstream out of the dashed box, where they will exert hydrodynamic and contact forces on neighbouring particles that eventually percolate the extruder and result in mass flow through the die. The flow rate through the extruder is therefore dependent on the magnitude of the constant force applied in the dashed box. In order to achieve steady flow over large displacements, particles are generated inside the dashed box at each timestep, such that the particle volume fraction (defined as the total volume of particles divided by the total volume) within this region remains at approximately $\phi = 0.5$. Note that there is no imposed volume fraction or particle number in the main body of the extruder: a constant volume fraction and driving force are maintained only within the dashed box. Downstream of this region the system dynamically reaches an equilibrium state with a driving-force-dependent flow rate, pressure drop and volume fraction. The downstream end of the geometry is open, so particles flowing out of the end of the die are removed from the simulation and subsequently omitted from contact force and stress calculations.

Simulations may be run ad infinitum, meaning time averaged as well as instantaneous flow features can be studied during steady flow. The imposed force magnitudes $F$ implemented in the present work are chosen such that the flow remains close to the hard particle limit, with $F/k_nd\ll 1$ (practically this means particle overlaps, quantified by $\delta$, remain $\ll0.01d$) and such that overdamping arising due to particle-fluid interactions is sufficiently thorough that particle inertia can be neglected. Conventional wisdom \cite{Hinch2011,Cates2014,Boyer2011} suggests that the rheology should consequently be Newtonian in that stresses scale linearly with shear rates, though it remains exquisitely sensitive to particle volume fraction. 
Significantly larger values of $F$ would lead to $\delta \to \mathcal{O}(d)$, a considerable violation of the approximately hard particle condition. The resultant soft particle rheology~\cite{Seth2011}, reminiscient of emulsions, for example, may be captured by the present model~\cite{Ness2015a} but is beyond the scope of the present study.
Moreover, large values of $F$ may further lead to high velocities and localised shear rates for which the Stokes number exceeds 1, meaning the flow might transition from viscous, quasi-Newtonian to inertial~\cite{Trulsson2012,Bagnold1954}. We maintain sufficiently small $F$ to avoid these two regimes.

For post-processing, we adopt a simple binning protocol to coarse grain the particle information, obtaining continuous
velocity and stress fields. Bins are generated with dimensions $d \times d \times 5d$ on a regular grid in the $x$--$y$ plane. The properties of each particle are accumulated in a bin if the coordinates of the centre of that particle lie in the bin. Particle velocities $\mathbf{v}_i$ are averaged within each bin, giving a coarse grained velocity vector $\mathbf{V}_k$, which has components $V_{k,x}$, $V_{k,y}$, $V_{k,z}$ where $k$ is the bin index. Per-particle stresslet contributions $\bm{\Sigma}_i$ are summed in each bin and divided by the bin volume $V_\mathrm{bin}$ to obtain the coarse grained stress tensor for each bin $\bm{\sigma}_k$, which has components $\sigma_{k,xx}$, $\sigma_{k,xy}$ etc. We thus obtain a coarse grained map of the velocity and stress profiles at each timestep which can be used to generate temporally-averaged profiles or to study transient features of the flow. The present study focusses on the steady state flow behaviour of the system; an analysis of the transient behaviour is deferred to future work.

\subsection{Simulation cases studied}

In the present study, we give a general description of the flow features predicted by the above model, followed by an overview of the roles of the model parameters. From the particle-fluid interaction, the available parameters are the magnitude of the fluid viscosity $\eta_f$ and the limiting surface separation constants $h_\text{min}$ and $h_\text{max}$. As discussed above, $h_\text{max}$ is chosen to be sufficiently large that all immediate neighbours will be considered when calculating the lubrication forces for a particle. It has been confirmed that setting $h_\text{max} = 0.1d_{ij}$ does not give significantly different rheological behaviour, though it does significantly increase the computational expense. Furthermore, it has been demonstrated recently \cite{Ness2016b} that the size of $h_\text{min}$ is important in determining the response of model pastes to minute deformations~\cite{Cates1998}, but that its influence during large paste deformations and in flows close to steady state is minor. Therefore, the primary fluid parameter available for investigation is the interstitial fluid viscosity $\eta_f$. 

The parameters present in the particle--particle interaction model are the particle stiffnesses $k_n$ and $k_t$ and the particle--particle and particle--wall friction coefficients $\mu_p$ and $\mu_w$ respectively. We take $k_t = (2/7)k_n$, following~\cite{Plimpton1995}. Each of the other parameters is varied independently, and the influence on flow behaviour is studied. A summary of each of the simulation Cases explored is given in Table~\ref{tab:cases}. Values of the particle--wall friction coefficient $\mu_w$ are kept low, to represent the very smooth extruder walls found in practice, while further wall resistance derives from the tangential component of the lubrication force acting between paste particles and constituent wall particles.
\begin{table}
\centering
\begin{tabular}{|x{1cm}|x{1.4cm}|x{1.2cm}|x{1.4cm}|x{1.4cm}|}
\hline
Case & Particle friction $\mu_p$ & Wall friction $\mu_\text{w}$ & Fluid viscosity $\eta_f$ & Particle stiffness $k_n$\\ \hline
A    & 1 & 0 & 0.05 & 10000\\
B   & 1 & 0.1 & 0.05 & 10000\\
C    & 0.1 & 0 & 0.05 & 10000\\
D   & 0.1 & 0.1 & 0.05 & 10000\\
E & 1 & 0 & 0.0005 & 10000\\
F & 1 & 0 & 0.05 & 20000\\ \hline
\end{tabular}
\caption{Summary of particle, fluid and wall properties investigated. $\eta_f$ and $k_n$ are given in simulation units $\rho d^2/t$ and $\rho d^3/t^2$ respectively, for time unit $t$, while $\mu_p$ and $\mu_w$ are dimensionless. In each Case, we set $k_t = (2/7)k_n$.}
\label{tab:cases}
\end{table}
As argued above, the paste flow regime investigated in this work is a hard-particle, viscous regime. It is therefore anticipated that, provided we remain in this regime, varying the parameters $k_n$ and $\eta_f$ will have a quantitative, but not qualitative, influence, since the relative paste viscosity, i.e. the paste viscosity divided by the fluid viscosity $\eta_f$, is independent of both when driving forces are scaled appropriately \cite{Chialvo2012,Ness2015a,Lin2015}.

\section{Predicted extrusion flow features}
\subsection{General description of flow}

\begin{figure*}
  \centering
\includegraphics[width=0.9\textwidth,valign=t]{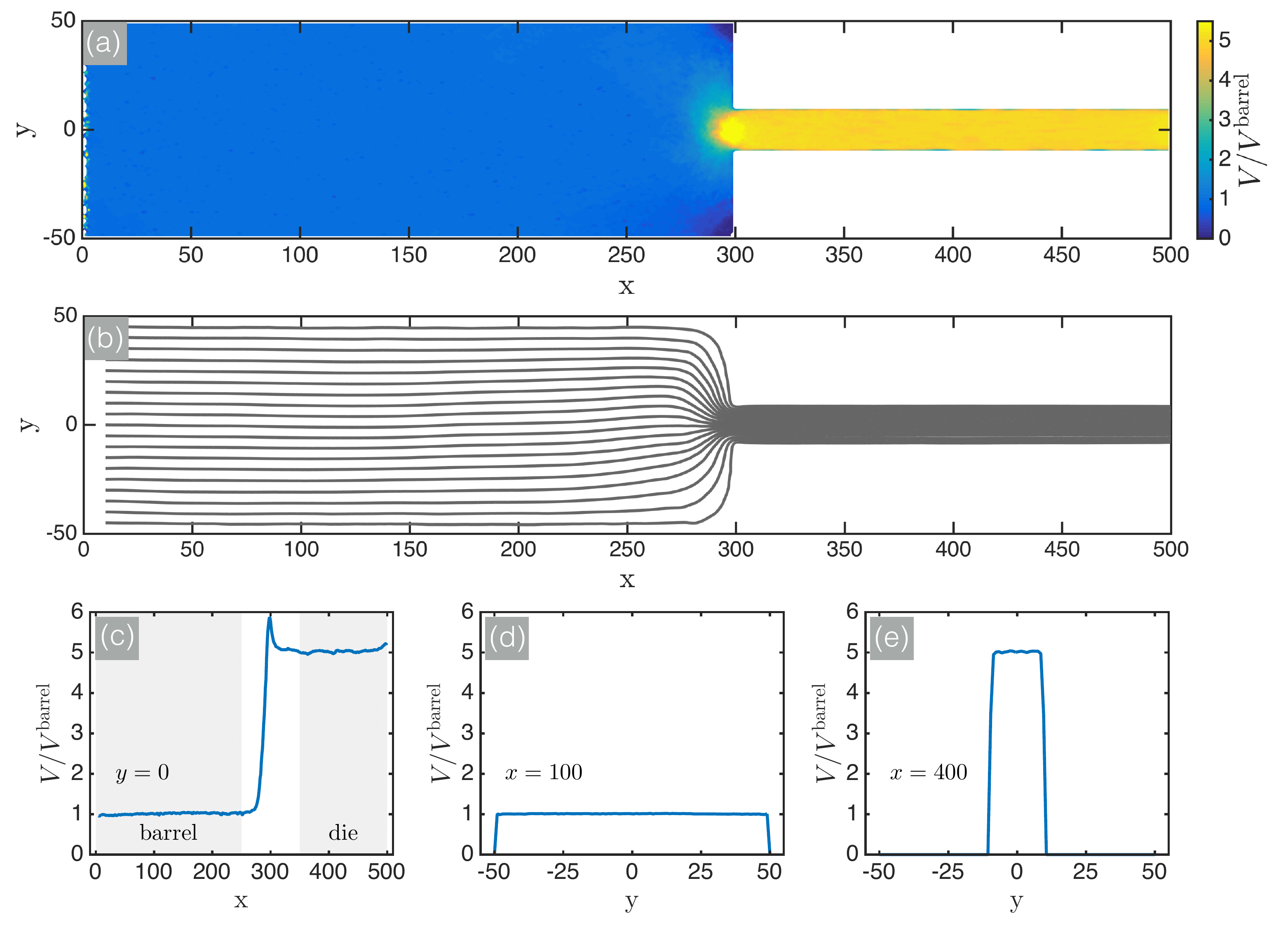}
\caption{
Velocity profile in the extruder for simulation Case A with $\Delta P = 5\times 10^5$Pa. (a) Velocity contour plot showing the total velocity magnitude $V$ scaled by the mean centreline downstream velocity in the barrel $V^\text{barrel}$; (b) Streamlines associated with (a); Velocity profiles along (c) $y=0$, showing the locations of the barrel and die; (d) $x=150$; (e) $x=400$.
}
\label{fig:velocity_contour}
\end{figure*}

For each Case, simulations are carried out for pressure drops $\Delta P$, set through the applied driving force $F$ (in units of $k_nd$), spanning two orders of magnitude. Five simulations are carried out for each value of $F$, each with different precise configurations of wall particles but representing the same large-scale geometry. The following results represent ensemble- and time-averages.

We first present a general picture of the flow features obtained using an initial set of parameters, Case A (Table~\ref{tab:cases}). 
Starting with an empty extruder, the paste flows in at the upstream end as particles are generated and flow is initiated in the dashed box, Figure~\ref{fig:geometry1}. There is no flow out of the die during this time, so the overall particle number increases steadily. Once the extrusion barrel fills up and a contact network~\cite{Peters2005,Majmudar2005} establishes between the barrel inlet and the die entry, particles begin to flow into, and through, the die to the outlet, eventually establishing a steady state. The particle number at this point is $\mathcal{O}(10^5)$. It is verified that a steady state has been reached by checking that the mass balance is satisfied such that the total number of particles present inside the extruder is approximately constant with time. Since the particle number is not fixed, the outlet flow minus the inlet flow fluctuates with time around zero.

\subsubsection{Extrusion velocity profiles}
\begin{figure*}
  \centering
  \includegraphics[width=0.7\textwidth,valign=t]{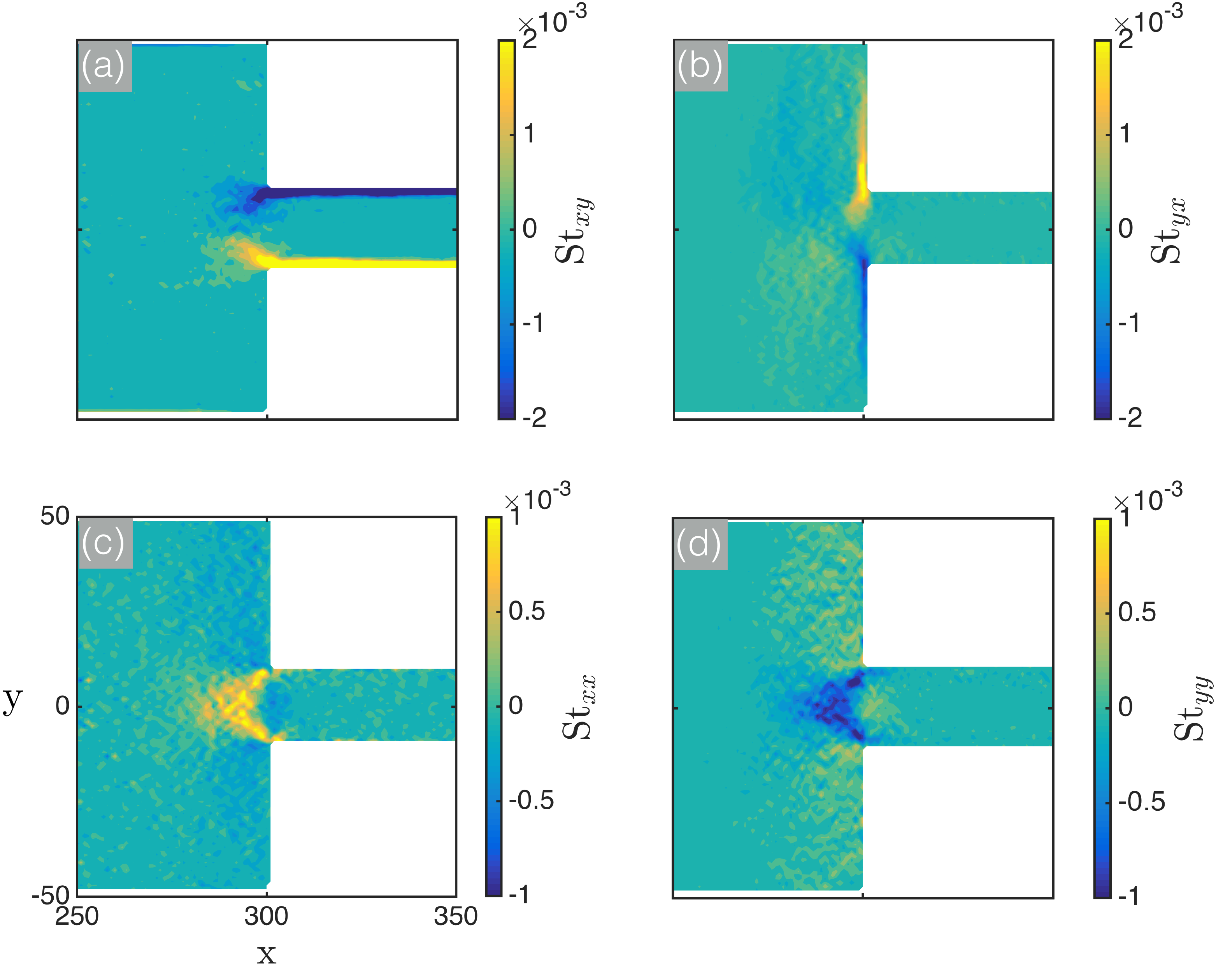}
     \caption{Velocity gradients at the die entry for simulation Case A with $\Delta P = 5\times 10^5$Pa, showing shear gradients (a) $dV_x/dy$; (b) $dV_y/dx$ and extensional gradients (c) $dV_x/dx$; (d) $dV_y/dy$. Velocity gradients are scaled with $\rho d^2/\eta_f$ such that the color scale represents Stokes numbers with respect to each flow gradient: (a) St$_{xy}$; (b) St$_{yx}$; (c) St$_{xx}$ and (d) St$_{yy}$. Stokes numbers remain $\ll1$ in all regions.
     }
     \label{fig:velocity_gradient}
\end{figure*}

A contour plot of the time-averaged velocity magnitude ($V = (V_{k,x}^2 + V_{k,y}^2 + V_{k,z}^2)^{1/2}$) profile is given in Figure~\ref{fig:velocity_contour}a, for $\Delta P = 5\times 10^5$Pa. For simplicity of interpretation, the velocity magnitude is scaled according to $V / V^\text{barrel}$, where $V^\text{barrel}$ is the mean centreline (i.e. at axial position $y=0$) velocity magnitude in the barrel. The associated streamlines are given in Figure~\ref{fig:velocity_contour}b, while velocity profiles along lines at $y=0$, $x=150$ and $x=400$ are given in Figures~\ref{fig:velocity_contour}c, d and e, respectively.

Together, the velocity profiles illustrate that there is viscous, uniform flow inside the extruder, that is qualitatively reminiscent of experimental velocity profiles measured using magnetic resonance imaging~\cite{Rabideau2012}, positron emission particle tracking~\cite{Wildman1999} (though accounting for the fact that they use a cone geometry so have avoided internal shear localisation), and theoretical work by Ref \citep{horrobin1999}. Specifically, it is observed that there is very little variation along the $x$-direction in the regions away from the die entry either in the barrel, between the inlet and around 10-20$d$ upstream of the die entry; or in the die, from around 5-10$d$ downstream of the die entry, to the outlet. To clarify this observation, the centreline velocity magnitude $V(y=0)$ is plotted in Figure~\ref{fig:velocity_contour}c. It is observed that the die velocity is approximately 5 times larger than that in the barrel, consistent with the aspect ratio of these regions, satisfying mass conservation.

In both barrel and die, the velocity is highly laminar and nearly independent of $y$-coordinate, indicating the presence of near-perfect plug flow throughout the extruder, away from the die entry, which we verify in Figures~\ref{fig:velocity_contour}d~and~\ref{fig:velocity_contour}e. The particle--wall friction coefficient $\mu_{w}$ is set to zero in this Case (see Table~\ref{tab:cases}), meaning any deviation from ideal plug flow, which would represent perfect wall slip, must be attributable to the inherent bumpiness of the boundary, that derives from its particle-based composition and offers some resistance to flow, supporting loads with non-zero $x$-components.
For the purposes of industrial paste processing, it is desirable to minimize the extent of shearing flow, thus moving the system as close to ideal plug-flow as possible. Perfect slip flow, often facilitated by a liquid slip layer at the boundaries, can lead to smoothly flowing extrudates with minimal surface defects, whereas shear flow close to the paste surface might result in instabilities being generated at the die outlet. Indeed, the present model, with its explicit value for $\mu_w$, may serve as a tool in future studies to investigate the role of wall friction on the formation  of shear zones close to extruder boundaries.
Note that the streamlines indicate a highly penetrative velocity field, where the flow extends almost completely into the corners of the die entry. Furthermore, even in this region there remains highly viscous flow, with no evidence of recirculating or unstable flow behaviour.
It therefore appears that the majority of the deformation, that is, the primary extensional and shearing part of the flow, is concentrated within a very narrow region close to the die entry.

To elucidate the deformation modes acting in this region, we present the velocity gradient in Figure~\ref{fig:velocity_gradient}, for shearing $\partial V_x/\partial y$, $\partial V_y/\partial x$ and extensional $\partial V_x/\partial x$, $\partial V_y/\partial y$ flows, noting that the $\partial/\partial z$ terms vanish. In each case, we give the velocity gradient scaled by $\rho d^2/\eta_f$, so that the numerical values act as representative Stokes numbers pertaining to the velocity gradient in the direction of interest: St$_{xy}$, St$_{yx}$, St$_{xx}$ and St$_{yy}$. The velocity gradients reveal considerable shear flow, both along the inner edges of the die, Figure~\ref{fig:velocity_gradient}a, extending to the outlet (not shown) and, surprisingly, along the vertical inner edges of the die entry, where there is considerable shearing flow in $\partial V_y/\partial x$, Figure~\ref{fig:velocity_gradient}b. The liquid phase migration paradigm suggests that pressure gradients culminating at the die entry corners lead to liquid migration away from this area. There is a consequent increase in particle concentration leading to jamming or solidification at these regions of the extruder. Instead, the present model suggests that a net flow of particles can be achieved along the inner edge of the die entry, allowing such pressure gradients to dissipate. We note that no significant spatial variations in the solids fraction are identified in the present study.

The flow at the entrance point to the die exhibits significant extension and compression, as indicated by the $\partial V_x/\partial x$ and $\partial V_y/\partial y$ profiles, Figure~\ref{fig:velocity_gradient}c~and~\ref{fig:velocity_gradient}d respectively. It is observed that the flow entering the die along the vertical extruder walls at the die entry is subject to compression, indicated by the squeezing of the streamlines at this point and the negative $\partial V_y/\partial y$. Simultaneously, the flow entering the die along the extruder axis (near to $y=0$) undergoes almost pure extension, as shown by the positive $\partial V_x/\partial x$, maintaining the incompressibility of the material so that a relative permeating liquid flow can be neglected. This stretching of the paste entering the die may be indicative of the above mentioned suction effect \cite{Mascia2006,Zhang2011,Patel2007} that is responsible for drawing liquid out of slower flowing die entry corners into the main extruding flow.
These velocity gradients serve to offer a thorough qualitative description of the main deformational flow at the die entry at steady state.

\begin{figure}[b]
  \centering
 {\includegraphics[width=0.4\textwidth]{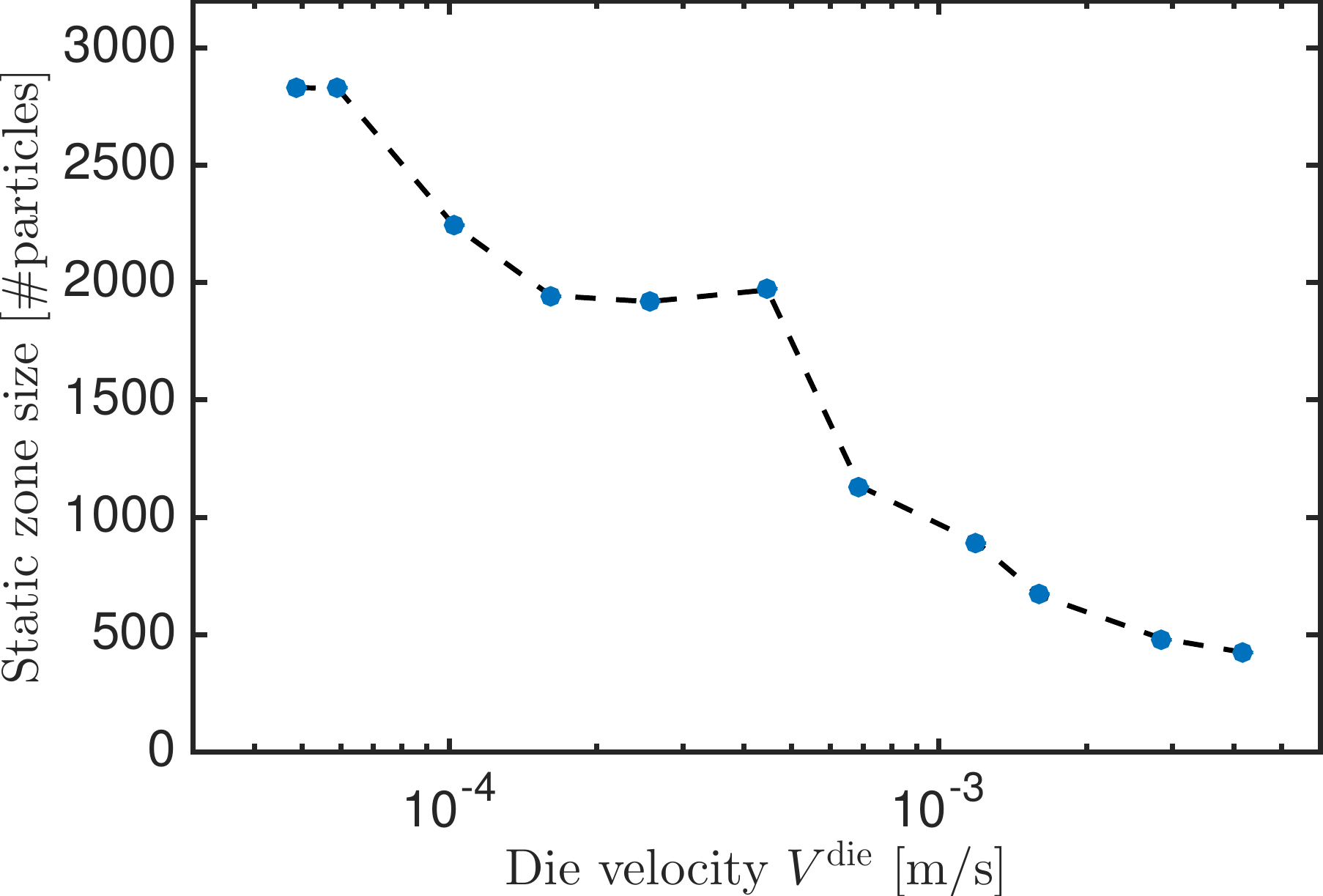}}
     \caption{Time-averaged static zone size for simulation Case A, counting all particles within the static zones, as a function of the die velocity $V^\mathrm{die}$.
     }
     \label{fig:velocity_static_zones}
\end{figure}

\begin{figure*}
  \centering
    \includegraphics[width=0.85\textwidth,valign=t]{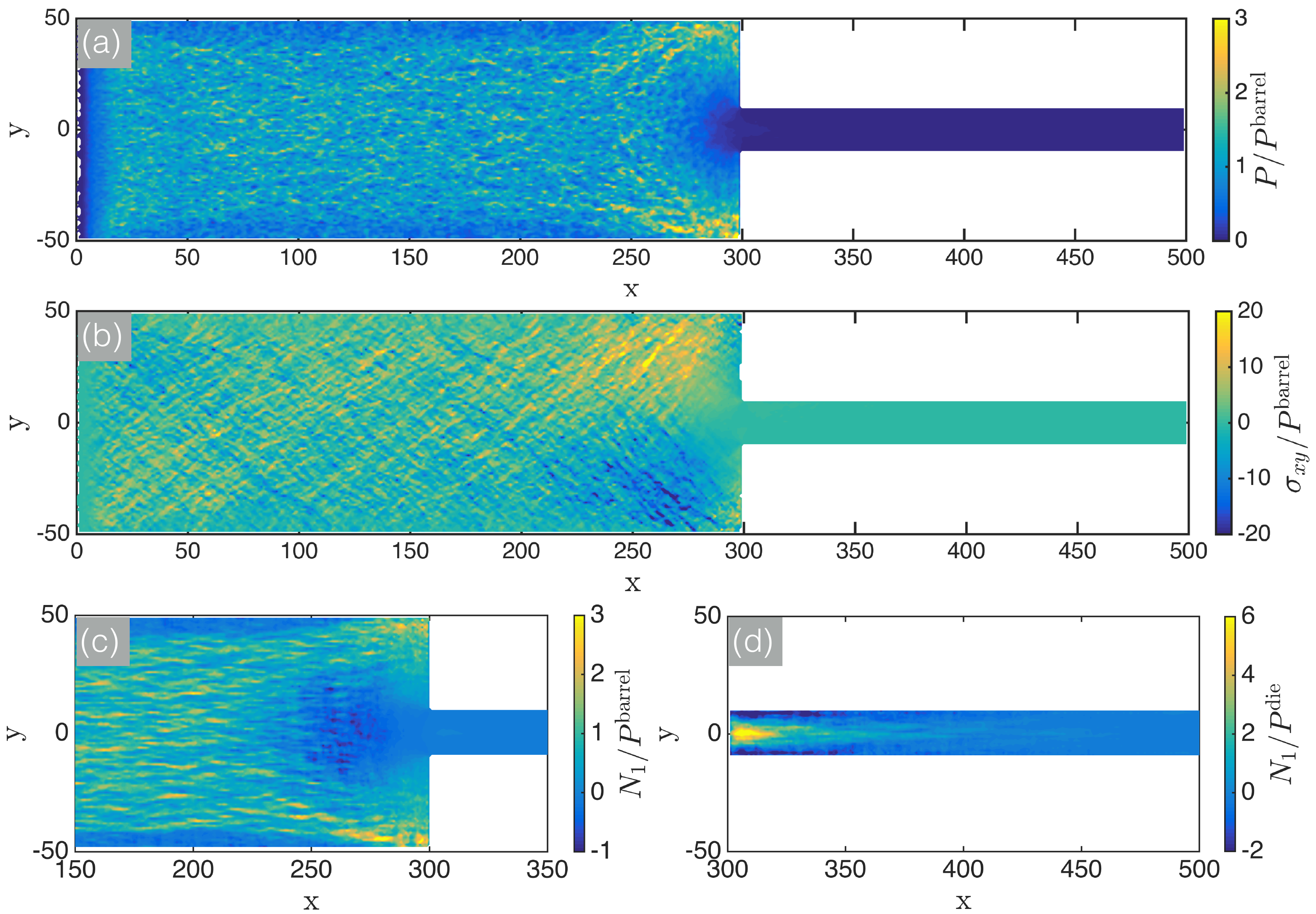}
     \caption{Internal stress profiles for simulation Case A with $\Delta P = 5\times 10^5$Pa.
     (a) Pressure $P$, scaled by the mean pressure at the upstream end ($25<x<50$) of the barrel, $P^\mathrm{barrel}$;
     (b) Shear stress $\sigma_{xy}$, scaled by $P^\mathrm{barrel}$;
     (c) First normal stress difference $N_1 = \sigma_{xx}-\sigma_{yy}$, scaled by $P^\mathrm{barrel}$;
     (d) First normal stress difference $N_1$, scaled by the mean pressure at the upstream end ($300<x<325$) of the die, $P^\mathrm{die}$.
     }
     \label{fig:stresses}
\end{figure*}

It is noted, particularly in Figure~\ref{fig:velocity_contour}, that the flowing regions do not extend fully to the bounding walls at $y = \pm 50$ for $x=300$, meaning there are small, nearly stagnant, regions in the corners of the die entry. For convenience, the extent of these stagnant regions near the die entrance is quantified by counting the number of coarse-grained cells for which the local velocity $V/V^\text{barrel}< 0.25$, and summing all the particles in those cells. The variation of stagnant region size with extrusion velocity is presented in Figure~\ref{fig:velocity_static_zones}. The simulation predicts that static zone size decreases with increasing mass flow rate. A static zone represents a region in which the solid contact network composed of paste particles is nearly stationary, while in principle the liquid phase may remain mobile by engaging in a porous-media type flow due to the osmotic pressure created by the flow in the central region \cite{Deboeuf2009} . For paste processing purposes, it is therefore desirable to minimize static region formation. The present results indicate a reduction of static zone size of up to 80\% when increasing the die velocity by an order of magnitude, suggesting that liquid phase migration deriving from such static zones may be reduced by operating at increased extrusion velocities. In practice, however, the risk of other instabilities is also introduced when changing the velocity~\cite{Benbow1993}, so factors other than the size of the static zone also have to be considered. The evolution of static zone size with increasing die velocity demonstrates that although the bulk $\Delta P \propto V^\mathrm{die}$ relationship holds (see later), the paste flow has to be treated as inhomogeneous and subtly rate-dependent in character.

\subsubsection{Extrusion stress profiles}

We next consider the steady state stress profile in the extruder, focussing on the pressure $P=\frac{1}{3}(\sigma_{xx}+\sigma_{yy}+\sigma_{zz})$, the shear stress $\sigma_{xy}$, and the first normal stress difference $N_1 = \sigma_{xx}-\sigma_{yy}$, presented in Figure~\ref{fig:stresses}. It is noted that erroneous (i.e. particularly low) stresses are found for $0 <x<10$ due to the boundary condition being implemented within this region, though the momentum equation remains satisfied when the imposed body forces are taken into account.

\paragraph{Pressure profile}
Consistent with the velocity profile, a rather uniform pressure profile is observed in the main flowing regions of the barrel for $50<x<200$, and the die for $350<x<450$, away from the die entry. A slight pressure reduction is observed within $10d$ from the barrel walls, where the limited shear flow is sufficient to perturb axially oriented contact chains that propagate from the die-entry corners, though the pressure remains largely uniform across the majority of the barrel width, with $\sigma_{yy}$ having zero gradient in $y$, to satisfy momentum conservation. The pressure behaves similarly in the die, with a largely uniform value across the width. 

It is found that the variation in $x$ within the barrel ($x<250$) is small, suggesting that, for the present Case of $\mu_w=0$, the 
pressure drop resulting from the inherent wall bumpiness is rather small. By contrast, a significant pressure drop, typically around an order of magnitude, is then observed at the die entry, in the region $250<x<300$. There is a further pressure drop, between one and two orders of magnitude though the absolute value is already considerably reduced following flow through the die entry, along the length of the die. The pressure then drops to zero at the outlet. This axial pressure drop is essentially what drives the net flow of paste through the extruder. 

A marked deviation from uniform pressure distribution is observed in the region $250<x<300$. An arching effect is observed, whereby regions of high pressure extend from just upstream of the die entry into the stagnant corner zones. The pressure reaches a maximum at these points. This arching pressure appears to be transmitted inhomogeneously through force chains that can be observed as narrow regions of high pressure protruding from near the corners of the square die entry region into the central part of the barrel. Such arching behaviour is reminiscient of simulations of gravity-driven silo discharges~\cite{Holst1999,Wang2015}, in which normal and tangential loads supported by frictional walls allow stable bridges to transiently form and collapse. For the nearly-frictionless walls considered in the present simulation, it is expected that the focussing of normal stresses on the square end walls at $x=300$ will be considerable, since only minimal $x$-direction loads can be supported by the barrel walls. This is consistent with practice, where the barrel wall is typically lubricated and would bear little load in $x$. There is a significant pressure drop in the $y$-direction at the die entry, for $275<x<300$. This radial pressure drop between the corners of the die entry and the die entry itself is responsible for the lateral flow of paste observed in this region.

\paragraph{Shear stress profile}
The shear stress contour plot in Figure~\ref{fig:stresses}b shows a profile that is comparable to that of the pressure. Notably, there are regions of significantly higher shear stress that correspond to the local arching effects. The shear stress in these regions is, consistent with the pressure, focussed into narrow bands, or chains, that propagate from the central point of the arch to close to the square die corners. The shear stress is approximately symmetrical in magnitude (though opposite in sign) about $y=0$. Averaging along the barrel, we find linear variation in $\sigma_{xy}$ between the wall values, but with significant fluctuations arising, particularly close to the die entry.

\paragraph{First normal stress difference}
The first normal stress difference $N_1$, Figure~\ref{fig:stresses}c, quantifies the relative magnitude of the compressive normal forces acting along the downstream ($x$) and radial ($y$) axes. The magnitude of $N_1$ is, significantly, comparable to the pressure, strikingly different from common non-Newtonian fluids. $N_1$ is positive ($\sigma_{xx}>\sigma_{yy}$) in the bulk region and largely follows the pressure across most of the extruder, indicating that the paste is compressed more in $x$ than in $y$, under the influence of the driving upstream force. Notably, however, there is a region of negative $N_1$ immediately upstream of the die entry. This suggests that the paste is compressed more in the $y$ direction in this region, consistent with the velocity gradient profile (Figure~\ref{fig:velocity_gradient}) which shows extensional flow (positive $\partial V_x/\partial x$) in this region. Such a normal stress difference would lead to die swell if the stress was relaxed immediately after the die entry without the die land. $N_1$ is also negative near the barrel walls due to the shearing motion in these regions. In Figure~\ref{fig:stresses}d, we rescale the contours to reveal the behaviour of $N_1$ in the die. We find that up to approximately 50 particle diameters downstream of the die entry, there is considerable inhomogeneity in $N_1$ across the die width, being negative near the walls and positive in the centre. The inhomogeneity is dissipated as flow proceeds downstream, reaching near homogeneity at the outlet. Such complex $N_1$ profiles are indicative of a propensity for paste deformation, suggesting that if the die were to be cut short at, say, $x=325d$, the outgoing flow could be highly unstable. The simulation therefore predicts that extrudate flows might be stabilised by using dies of sufficient length to allow $N_1$ variations to dissipate. 

\begin{figure*}
  \centering
   {\includegraphics[trim = 62mm 30mm 34mm 45mm, clip, width=150mm]{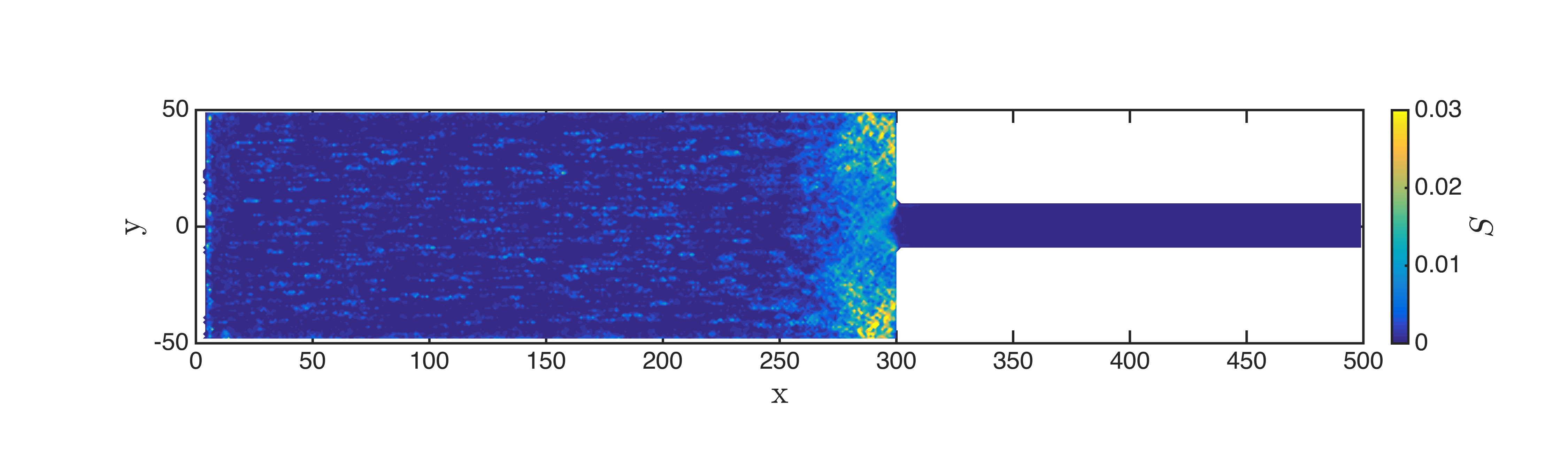}}
     \caption{Rate of energy dissipation per unit volume for simulation Case A, calculated as $S = \bm{\sigma}\colon\bm{\dot{\gamma}}$, where $S$ has units of $\rho d^2/t^3$.
     }
     \label{fig:dissipation_contour}
\end{figure*}
Though the stress data provide insight into the nature of the flow within the extruder, a more complete description of the origin of the extrusion pressure drop may be obtained by examining the localised rates of energy dissipation. We obtain a coarse grained energy dissipation rate profile $S$ using the coarse grained stress $\bm{\sigma}$ and strain rate field $\bm{\dot{\gamma}}$ (obtained from the velocity gradients) according to $S = \bm{\sigma}\colon\bm{\dot{\gamma}}$. We find that the majority of energy dissipation is heavily focussed around the die entry region, with particular emphasis at the corners of the square entry. This result reveals an interesting interplay between the rates of deformation and the stresses near the die entry. Specifically, we note that the maximal rates of dissipation occur at the extreme corners of the square die entry, where the paste is almost stationary as quantified earlier as static regions. Furthermore, it is demonstrated that the regions of highest shear stress, which propagate in narrow bands or chains from near the die corners into the centre of the barrel, do not in fact correspond to regions of significant energy dissipation. Conversely, the low pressure region at the point of die entry, where flow is dominated by extension in $x$, has a relatively high level of energy dissipation. Together, these results suggest that the most important flow behaviour contributing to the dissipation within the extruder comes from the jamming and nearly arrested flow in the barrel corners, as well as the extensional flow at the die entry. These results provide directly usable guidelines for extruder design to minimise energy dissipation, and hence pressure drop.

\subsection{Linking particle properties to extrusion behaviour}

The bulk flow behaviour of the paste with respect to the previously identified model parameters is finally assessed by considering the relationship between the velocity in the die during extrusion and the extrusion pressure drop. Here, we show the full advantage of our discrete element simulations, by demonstrating how the predicted bulk flow behaviour responds to changes in the particle-scale properties.

\begin{figure}
  \centering
      \includegraphics[width=0.495\textwidth,valign=t]{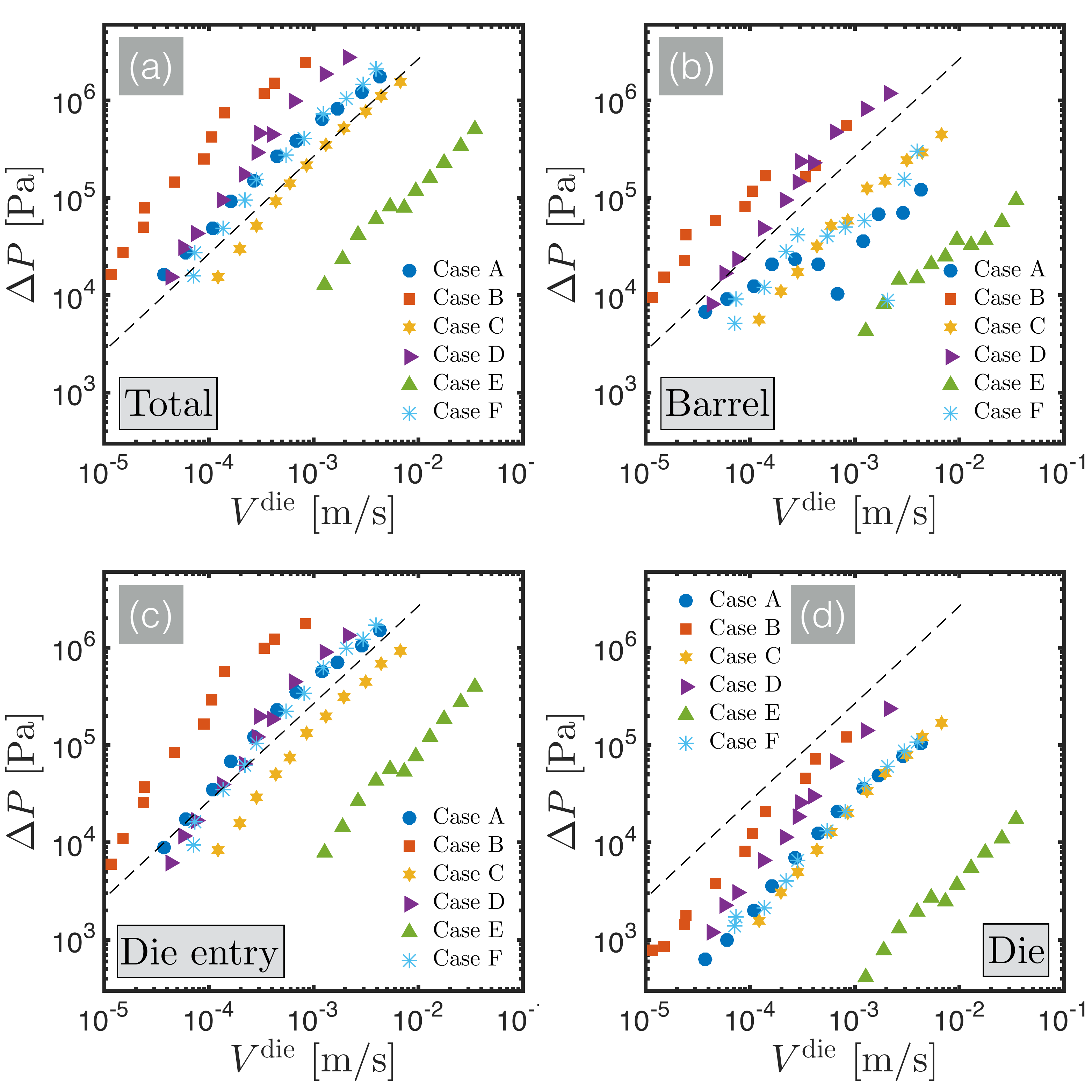}
     \caption{Pressure drop versus die velocity for simulation Cases~A~-~F, as described in Table~\ref{tab:cases}.
     (a) Total pressure drop across extruder;
     (b) Barrel pressure drop;
     (c) Die-entry pressure drop;
     (d) Die pressure drop. Dashed black line in each figure indicates a gradient of unity.
     }
     \label{fig:pressure_drops}
\end{figure}

The pressure drop is found to vary approximately linearly with extrudate velocity, for the entire range of velocities considered, Figure~\ref{fig:pressure_drops}a. Such behaviour may be related to the nature of the lubrication contact model described in Section~\ref{sec:contact_model}, which, assuming sufficiently hard particles and particle Stokes numbers considerably less than unity (which we verified in Figure~\ref{fig:velocity_gradient}), leads to suspension stresses that scale linearly with shear rates under steady state shear flow~\cite{Boyer2011,Ness2015a}. Notably, however, the linear scaling does not, necessarily, imply that the rheology of the material is Newtonian. Specifically, phenomena such as non-locality~\cite{Kamrin2014,Kamrin2015}, localised volume fraction variations and considerable normal stresses may still arise, leading to a highly complex relationship between the supposed viscosity of the material and the flow circumstances.

It is observed that no apparent bulk yield-stress behaviour is predicted by the present model. This can be explained by the boundary conditions of the simulation, in which the global volume fraction within the extruder is not fixed. Therefore, for any applied force at the barrel inlet, the volume fraction within the extruder evolves over time and eventually adopts a value such that flow may occur. Other than an oscillating boundary effect that is consistent with well documented previous works~\cite{Louge1994,Besseling2009}, the volume fraction throughout the extruder is found to be approximately uniform, with a value that lies marginally below the respective critical volume fraction for jamming~\cite{Liu1998} for the relevant set of model parameters. It is widely acknowledged that the critical volume fraction for jamming is crucially dependent on the particle-particle friction coefficient~\cite{DaCruz2005,Sun2011,Guy2015}. Simulations in a comparable geometry \footnote{Unpublished simulation results using a similar contact model but based on a channel flow geometry (C. Ness and J. Sun)} indicate that a precompaction process, whereby the paste volume fraction is forced to be above its critical value for jamming or flow arrest, introduces yield stress behaviour into the material. Thus, the present model may predict yield stresses that arise due to strong confinement effects, but this, and other sources of yield stress behaviour such as particle-particle cohesion~\cite{Luding2008}, are deferred to future investigations.  

The variation of extrusion pressure drop $\Delta P$ with die velocity is given in Figure~\ref{fig:pressure_drops}, for each of the simulation Cases A~--~F defined previously. We present the total pressure drop, the pressure drop in the barrel, the pressure drop due to paste deformation at the die entry, and the pressure drop in the die. For each pressure drop, results are presented for each of simulation Cases~A~--~F. In addition, a dashed line is included to indicate a gradient of unity. The present results exhibit linear scaling of each pressure drop with respect to extrusion velocity, demonstrating that our model predicts the 4- (as opposed to 6-) parameter Benbow-Bridgwater equation (Equation~\ref{eq:1}) is suitable for pastes of repulsive (nearly-) hard spheres in the viscous regime.

Consistent with the previous discussion regarding the independence of rheology on particle stiffness $k_n$ provided the hard-particle limit is adhered to, it is observed in Figure~\ref{fig:pressure_drops} that the results for Cases~A~and~F, corresponding to particle stiffnesses $k_n$ and $2k_n$ respectively, are nearly identical. This finding reveals that the extrusion rheology predicted by the present model is independent of particle stiffness, while $F$ remains $\ll k_nd$. Recent experiments using emulsions have illustrated how soft particles might respond in similar flow scenarios~\cite{Hong2015}, suggesting a further future extension of the present model.
Comparing Cases~A~and~E, a dramatic increase in extrusion flow rate is observed for reduced suspending fluid viscosity, at fixed $\Delta P$. It is noted, however, that this increase in flow rate is approximately two orders of magnitude, for a reduction in $\eta_f$ of the same proportion. Assuming that local shear rates scale directly with the overall extrusion flow rate and considering the paste flow behaviour in terms of its viscous number $\eta_f \dot{\gamma}/P$~\cite{Jop2006,Forterre2008,Lemaitre2009,Boyer2011}, applicable for hard-particle and entirely non-inertial flows, therefore, it is determined that Cases~A~and~E represent comparable flow states, as the viscous number is approximately constant while the local shearing Stokes numbers remain $\ll1$.

The roles of particle-particle and particle-wall friction are illustrated in Cases A~to~D. Total pressure drop predictions (Figure~\ref{fig:pressure_drops}a) are in line with expectation: increasing particle-particle friction (compare Cases A, $\mu_p =1 $ and C, $\mu_p =0.1$) increases total pressure drop; increasing particle-wall friction (compare Cases A and B  or Cases C and D, $\mu_w = 0 $ and $\mu_w =0.1$ respectively) increases total pressure drop. Comparing the relative barrel, die-entry and die pressure drops is, however, more revealing. For flow within the barrel, Figure~\ref{fig:pressure_drops}b, the pressure drop is found to be remarkably insensitive to the particle-particle friction, illustrated by comparison between Cases~A~and~C or Cases~B~and~D. Barrel flow is, however, highly sensitive to particle-wall friction, indicating that the entirety of the dissipation within this region derives from particle-wall interactions, regardless of the frictional properties of the particles. Analogous behaviour is observed for the die pressure drop, Figure~\ref{fig:pressure_drops}d, though the pressure drop magnitudes are considerably lower than those in the barrel. That the die pressure drop is insensitive to particle-particle friction but highly sensitive to particle-wall friction has important consequences for Benbow-Bridgwater modelling. Specifically, it may be concluded from the present results that the $\beta$ parameter in Equation~\ref{eq:1} is relatively independent of the surface properties of the constituent paste particles, but is highly sensitive to the processing conditions, and linked to the interaction between the paste and the extruder walls that we quantify here using $\mu_w$.

The pressure drop at the die-entry dominates the total pressure drop within the extruder. Comparing Cases~A~through~D reveals that the pressure drop magnitude is sensitive to both the particle-particle $\mu_p$ and particle-wall $\mu_w$ friction coefficients. From the velocity profile and velocity gradient contour plots in Figures~\ref{fig:velocity_contour}~and~\ref{fig:velocity_gradient} respectively, it can be deduced that the pressure drop at the die entry derives from both a predominantly elongational paste deformation across $275<x<300$, and a lateral ($y$-direction) flow of material along the bounding wall at $x=300$. The former may be dominated by details of particle-particle interaction, whereas the latter may be more sensitive to the particle-wall interaction. 
This dependence implies that the extrusion pressure drop is governed both by the material properties and the processing condition, suggesting a more complex relationship between $\alpha$ and the paste and paste--wall interactions, quantified here as $\mu_p$ and $\mu_w$.

Our model, therefore, provides a useful basis for obtaining bulk extrusion parameters, for example in the context of the Benbow-Bridgwater equation, that arise directly from specified particle-particle and particle-wall interactions.
It is anticipated that the model and the present findings may serve as a framework for establishing links between particle details and paste flow behaviour during extrusion, useful for model development as well as for aiding predictive future paste formulation and process design.

\section{Concluding remarks}
Using discrete element method simulations of paste extrusion through a square-entry die, we have revealed detailed deformation and stress features, which are difficult to obtain from either experiments or continuum simulations. In particular, combined shear and extensional flow near the die entry is shown through the velocity gradient field; patterns of the normal stress difference are linked to the potential for die swell and flow instabilities at the extruder outlet; energy dissipation is shown to be significant in regions where either stress or extensional deformation rate is high. These insights are useful for practical control or design of paste extrusion processes. Furthermore, our results help to expose the inherently discrete nature of the paste and inhomogeneity of the flow in extruder, drawing on rheological phenomena that originate at the particle-particle interaction level. 
The model itself may be used directly by industrialists as a tool to inform design and optimisation of extrusion processes, since its implementation is considerably less computationally expensive than alternative multiphase approaches that resolve individual particles and fluid motion. Moreover, the present findings may serve as a basis for future continuum descriptions of extrusion flow that take full account of the microscale physics that underlie the rheology.

The present model and results may be built upon and utilised in future in a number of ways, in order to strengthen and extend the findings given here.
First, the model does not explicitly conserve the mass of the fluid. This means that in regions of volume fraction gradient (which, as we point out above, are very small or negligible in the present case) it is assumed that the fluid may permeate the solid matrix at infinitely high rate. Further work is necessary to test the effect of time-dependent fluid pressure dissipation due to finite permeability during extrusion. 
Another limitation of the model is its inability to capture free surface flows. Specifically, we do not capture the flow behaviour of the paste when it exits the extruder. This is a crucial region of flow, where the engineer typically determines the outcome and success of the extrusion process, i.e. smooth, flowing extrudate or unstable, inhomogeneous, poorly mixed paste. In the present model, we have used the normal stress difference within the die as an indicator of such defects.
Based as it is on a classical discrete element method algorithm~\cite{Cundall1979}, numerous extensions may be made to our model to tailor it to specific pastes or materials of interest. For example, it may be directly extended to account for wide particle size distributions~\cite{Gu2016}, more complex particle-particle interactions such as adhesion~\cite{Gu2014} or electrostatic repulsion~\cite{Mari2014}, and non-spherical particle shapes~\cite{Estrada2008,Estrada2011}. In each case, the present model provides a basis for predicting the influence on the overall bulk rheology deriving from these particle details.

Overall, our model and findings have elucidated links between microscale physics and bulk extrusion behaviour, providing a foundation for the development of future continuum models as well as guiding future industrial practice for formulation and design. This can have broad impact on paste processing, and more widely in soft matter rheology across industry.
\section{Acknowledgements}
This work was funded by the UK Engineering and Physical Sciences Research Council and Johnson Matthey through a CASE studentship award, and benefitted from discussions with B.M. Guy, M. Hermes and W.C.K. Poon.

\bibliography{library}
\appendix

\end{document}